# Skyrmion strings and the anomalous Hall effect in $CrO_2$

H. Yanagihara* and M.B. Salamon†

*Department of Physics and Materials Research Laboratory,*

*University of Illinois at Urbana-Champaign*

*1110 W. Green Street, Urbana IL 61801*

## Abstract

As in such 2D systems as the XY model, topological (or singularity point) defects are thought to play a crucial role in the phase transitions of 3D spin systems. In double exchange (DE) ferromagnets, the conduction electrons are strongly coupled with core spins through Hund's rule, and in the presence of non-trivial spin texture, acquire a Berry phase contribution to the anomalous Hall effect (AHE). We combine Hall and magnetization data on $CrO_2$ with a thermodynamical scaling hypothesis to confirm that the critical behavior of the topological-spin-defect density is consistent with that of the heat capacity. This analysis is the first experimental confirmation of the topological character of critical fluctuations.

*Electronic address: yanagiha@uiuc.edu

†Electronic address: salamon@uiuc.edu

Topologically stable defects in materials often play a key role in phase transition phenomena. Topological defects (or vortices) are known to drive phase transitions in a large class of 2D systems, including superfluid films, 2D crystals, and planar-spin (XY) models [1, 2]. In 2D systems, singular excitations (topological defects) can be separated from nonsingular ones (spin waves) and treated analytically, leading to the conclusion that topological defects are both necessary and sufficient to describe the critical properties. Because the role of topological defects in 3D systems has not been investigated analytically, we must rely on Monte Carlo (MC) calculations[3–5] for evidence that topological defects play a crucial role in the phase transitions of, e.g. 3D Heisenberg ferromagnets. In 1989, Lau and Dasgupta (LD)[4] performed MC calculations on the classical 3D Heisenberg model, both allowing and suppressing topological defects. Without such defects, the system remains ordered at all temperatures, indicating that topological excitations play a central role in the phase transition. Allowing topological defects, they concluded that $d\langle n\rangle/dT \sim |t|^{-\psi}$, where $\langle n\rangle$ is the defect density, $t = (T - T_C)/T_C$, $\psi \sim 0.65$ and suggested that $\langle n\rangle$ is a *disorder* parameter. Several years later, however, Holm and Janke (HJ)[3] repeated the MC simulation using larger lattices. They reasoned that quasi-local defects should have energy-like character and, contrary to LD's conclusion, found *weak evidence* that $d\langle n\rangle/dT \sim |t|^{-\alpha}$; i.e., controlled by the heat-capacity exponent $\alpha$. While the MC evidence is strong, it is very important to establish the existence of such defects experimentally and to study how they are related to the ferromagnetic-paramagnetic phase transition.

The topological defects detected by LD produce a non-trivial spin texture that, in the presence of strong coupling between the conduction-electron and core spins, contributes to the Hall effect via the induced Berry phase [6–10]. As this contribution requires the presence of non-zero magnetization (and spin-orbit coupling) we denote it here as "the anomalous Hall effect (AHE)," even though it is not simply proportional to the magnetization. We assert that in double-exchange (DE) systems the strong coupling between the spin of the atomic core and that of the mobile electron make the AHE a sensitive probe of the nature of topological spin defects in 3D systems.

According to recent insights into the band structure of $CrO_2$ by Korotin et al.[11], $CrO_2$ can be regarded as a *self-doped* DE system due to strong hybridization of $d_{yz}$ and $d_{zx}$ orbitals with oxygen $p$ orbitals and Hund's-rule coupling of these itinerant $d$-electrons with the localized $d_{xy}$ core. With one $d$-electron localized and one itinerant $Cr^{4+}$ is similar to Mn ions in some doped manganites that show colossal magnetoresistance (CMR). Unlike these CMR manganites, $CrO_2$ has no strong Jahn-Teller effects, no disorder and no formally different valence [11]. Except for the origin of ferromagnetism, $CrO_2$ is more like a conventional ferromagnetic metal such as Ni and therefore can be an appropriate material in which to study *pure* DE ferromagnet properties.

$CrO_2$ films were grown by ambient pressure chemical vapor deposition on rutile (bct structure) $TiO_2$ (100) substrates following the recipe developed by Ishibashi et al.[12] The saturation moment of this material at low temperature is close to 2 $\mu_B$/f.u., suggesting that the sample is stoichiometric. The sample reported here is $\sim$ 2300 Å thick and was patterned by photolithography and Ar-ion beam milling. In order to measure resistance and Hall voltages along the inequivalent in-plane directions, the sample was patterned as shown in the lower inset of Fig. 1; that is, the current path is "L" shaped, the bars aligned along $b(\langle 010\rangle)$and $c-$axes ($\langle 001\rangle$) respectively. The width of the current path is 100 $\mu$m and the distance between two voltage leads for the resistance measurement is 500 $\mu$m. The structural quality of the film was checked by x-ray diffraction; we could find no chromium oxide signals besides the correct phase of $CrO_2$(200). The off-specular reflections show 2-fold symmetry confirming epitaxial growth. The residual resistivity ratio ($RRR \equiv \rho_{300K}/\rho_{10K}$) is another criterion of quality for a metallic sample and is about 66 for $c-$axis and 20 for $b-$axis (Fig. 1), similar to earlier results [13]. The resistivity in both directions shows an inflection point around $T$ = 380 K corresponding to $T_C$. As shown in the inset of Fig. 1, using the Fisher-Langer model for $d\rho/dT$ [14], we determined $T_C$ = 383.8 K and critical exponent $\alpha$= -0.17; the latter agrees with the value from the scaling relation $\alpha + 2\beta + \gamma = 2$, with $\beta = 0.37, \gamma = 1.43$ for $CrO_2$[15]. This agreement of the Fisher-Langer exponent with critical exponents from the magnetization measurements confirm that $CrO_2$ is a 3D Heisenberg-like ferromagnet. We note that a previous scaling analysis of a polycrystalline sample gave totally different exponent ($\alpha = -0.37$)[16]. It appears that the Fisher-Langer relation is valid for this class of ferromagnets despite the unconventional ferromagnetic mechanism and relatively low (but metallic) conductivity.

In Fig. 2, we plot the field dependent Hall resistivity at various temperatures from 10 to 480 K. The Hall resistivity of magnetic materials is usually expressed by

$$\rho_{xy} \equiv E_{xy}/j_x = R_0 B + R_S \mu_0 M, \qquad (1)$$

with $R_0$ the ordinary and $R_S$ the anomalous Hall coefficients. Throughout our experiment, magnetic fields were applied perpendicular to the film plane, $(H//\langle 100\rangle)$ and therefore the demagnetization factor can be regarded as unity. As one can see, both directions of the Hall resistivity (open circles and lines) are identical within an experimental error, in agreement with Onsager's principle that $\rho_{xy} = -\rho_{yx}$ regardless of crystal orientation. As noted above, we term $\rho_{xy}^A = \rho_{xy} - R_0 B$ the anomalous Hall resistivity, or AHE. Generally, the low temperature Hall effect arises only from the ordinary (OHE) if there is no intrinsic spin disorder or frustration [9]. The OHE is hole-like at low temperatures, in agreement with earlier work [13] but appears to have a kink in the low temperature $\rho_{xy}$ curves at around $B = 1$ T, probably not due to AHE but rather OHE from the complicated multiband structure[17]. Above 100 K, the AHE gradually arises and the $\rho_{xy}$ curves become large and negative, indicative of an electron-like AHE. With increasing temperature, the magnitude of $\rho_{xy}$ maximizes just below $T_C$(=384 K) with our achievable field, and gradually decreases. This behavior is qualitatively quite similar to that of CMR systems and should be explained by the Berry phase theory [6, 10] rather than conventional spin scattering mechanisms such as side jump and skew scattering[18].

According to the MC simulations[3–6], topological defects are dilute at low temperatures with positive $Q = +1$(hedgehogs) and negative $Q = -1$(anti-hedgehogs) separated by the lattice spacing $a$ and strongly bound into pairs (or strings). Each pair acts as a lattice-spacing-size magnetic dipole. The temperature dependence of the topological spin-defect pair density $n$ at low temperatures is given by $\langle n\rangle \propto \exp(-E_C/k_B T)$, where $E_C$ is the core energy of a pair. Well below $T_C$, the defect pairs are dilute and behave independently. When spin-orbit coupling $\lambda_{so}$ is taken into account, a gauge field $\langle b_z\rangle$ appears due to a slight imbalance between the numbers of pair dipoles oriented parallel and antiparallel to magnetization $M$:

$$\begin{aligned} \langle b_z\rangle &= -\frac{\Phi_0}{2\pi a^2}(\langle n_+\rangle - \langle n_-\rangle) \\ &= -\frac{\Phi_0}{\pi}\frac{\langle n\rangle}{a^2}\sinh(\frac{\lambda_{so} n_{el} M a^3}{k_B T g \mu_B}). \end{aligned} \tag{2}$$

Here, $\Phi_0 \equiv hc/e = 4.1357\times 10^{-15}$ Wb, and $n_{el}(\approx 0.5)$ is the number of carriers per unit cell. We neglect higher topological charge defects ($|Q| \geq 2$). Because $\lambda_{so}$ is small compared to $k_B T$, we can expand the hyperbolic function to find $\langle b_z\rangle \propto M$, a result confirmed over wide range of temperatures by a MC simulation of the double-exchange model by Calderón and Brey. [6]. This $M$-dependent gauge field produces the AHE. Combining the data in Fig. 2 with $M - H$ curves with $H//[100]$, we separated $R_S$ and $R_0$ by fitting to Eq. (1). The temperature dependent $R_S$ is shown in Fig. 3. Qualitatively, this result shows very good agreement with Ye et al.'s prediction as well as Calderón and Brey's calculation; that is, $R_S$ grows exponentially at low temperature, has an inflection point near $T_C$, and then shows a maximum at $\sim T_C + 30$ K before decreasing gradually at high temperature. Expanding Eq.(2) in the low temperature regime, we can simply describe $R_S$ by

$$R_S \equiv \frac{\rho_{xy}^A}{\mu_0 M} = -\frac{\Phi_0}{\pi}\frac{\lambda_{so}}{k_B T}\frac{a n_{el}}{g\mu_B\mu_0}\langle n\rangle R_0 \propto \exp(-E_C/k_B T)/T. \tag{3}$$

We fit the data shown in Fig. 3 by using Eq. (3) over a temperature region where the independent string picture should be valid, obtaining an exponential increase with $E_C \sim 1100$ K $\approx 3T_C$. Monte Carlo calculations for DE or Heisenberg models give different values for $E_C$, but agree on the topological defect pair density at $T_C$, $\langle n\rangle|_{T=T_C} \sim 0.05$ [3, 4, 6], fixing the exponential prefactor. With $n_{el} \approx 0.5$, we obtain a large, but still plausible, value for the spin-orbit coupling $\lambda_{so} \approx 30$ K. Our $E_C$ is substantially smaller than expected from combining Ostlund's [19] analytic value $4\pi J$ for the Heisenberg model with the MC critical temperature $T_C = 1.5J$ to get $E_C = 8.4T_C$; it is also smaller than obtained by the MC simulation of a classical double-exchange ferromagnet ($E_C \sim 7T_C$). [6] However, the energy for small deviations in the angle between neighboring spins is proportional $1 - \cos\theta \approx \theta^2/2$ in the Heisenberg case, while the double-exchange energy varies as $1 - \cos(\theta/2) \approx \theta^2/8$, decreasing the core energy to $\pi J$ for a double exchange system [20]. Using the transition temperature $T_C = 1.06J$ from Calderón and Brey, we see that $E_C = 3T_C$ is the expected core energy.

The MC simulations show that the exponential rise in $\langle n\rangle$ gives way to an inflection point near $T_C$, causing $d\langle n\rangle/dT$, and therefore $dR_S/dT$, to exhibit a peak. Lau and Dasgupta indicated that the peak behaves $\sim |t|^{-\psi}$, with $\psi \approx 0.65$, while HJ's simulation suggested the weaker singularity $\sim |t|^{-\alpha}$, where $\alpha$ is the specific heat exponent. Ye, et al. endorsed HJ's suggestion, but expected this weak anomaly to be obscured by the finite field required to observe the AHE[10]. We do see an inflection-like anomaly at $T_C$ but cannot extract the critical behavior of the topological defects from Fig. 3. Rather, we extend HJ's scaling ansatz by including the scaled magnetic field and seek to collapse the Hall data as commonly done for other properties near a critical point. Scaling in favor of the magnetic field by setting $L^{\beta\delta/\nu}h = 1$, we obtain

$$\begin{aligned} \langle n \rangle &= \langle n \rangle^{reg} - L^{(\alpha-1)/\nu} f(L^{1/\nu}t, L^{\beta\delta/\nu}h) \\ &= \langle n \rangle^{reg} - h^{(1-\alpha)/\beta\delta} f(t/h^{1/\beta\delta}, 1), \end{aligned} \quad (4)$$

where $t$ and $h$ are reduced $T$ and $H$, respectively, $f(x,y)$ is a scaling function of the free energy [21] and $\langle n \rangle^{reg}$ is the non-critical part of the defect-pair density. Because $\rho^A_{xy}$ is proportional to the magnetization, we eliminate the explicit field dependence in Eq.(4) via the usual scaling behavior of the magnetization, $m(t,h) = h^{1/\delta} g(t/h^{1/\beta\delta}, 1)$ to obtain the following scaling equation for $\langle n \rangle$,

$$\langle n \rangle = \langle n \rangle^{reg} - m^{(1-\alpha)/\beta} \frac{f(x,1)}{g(x,1)^{\delta}}, \quad (5)$$

where the scaled temperature is $x = t/h^{1/\beta\delta}$. Combining this with Eq.(3), we can express the anomalous Hall resistivity as a function of $m$ and scaled temperature,

$$\rho^A_{xy} T = \rho^0_{xy} T_C m (1 - D(x) m^{(1-\alpha)/\beta}). \quad (6)$$

In this equation, $D(x)$ is the fraction term in Eq. (5) divided by $\langle n \rangle^{reg}$ and $\rho^0_{xy}$ contains all other constants. Fig. 4 shows the various isotherms of $\rho_{xy} T$ as a function of the reduced magnetization, $m \equiv M/M_0$. At low temperature, both $m$ and $\rho_{xy}$ increase rapidly as the domains are swept out. The slight positive increase in the Hall resistivity close to saturation is from the OHE. As temperature increases, the anomalous term dominates and the signature of the ordinary term becomes difficult to detect. Close to the $T_C = 384$ K, the $\rho_{xy} T$ curves reach a maximum at $m \approx 0.6$. The extremum of Eq. (6) is at $m = (\beta/(D(1-\alpha+\beta)))^{\beta/(1-\alpha)} \approx (0.24/D)^{0.316}$ for $\alpha = -0.17$ and $\beta = 0.37$. The curve of Fig. 4 follows the experimental data near the critical temperature for $D(x=0) = 1.2$. The dashed curve represents the best fit when the LD exponent $\psi = 0.65$ is used in place of $\alpha$; $D = 0.8$ in that case. Along $x = 0$, we expect $\rho^A_{xy} \to 0$ as $m \to 1$ so that $D(0) \approx 1$ is not surprising. It is remarkable that the wide range of fields and temperatures that comprise the plots in Fig. 4 collapse as they do. This result strongly supports HJ's suggestion that $d\langle n \rangle/dT$ has the same exponent as the heat capacity and adds credance to LD's assertion that the topological spin defects play crucial role in the phase transition, .

We conclude that the AHE of $CrO_2$, a *self-doped* DE system, is consistent with Ye et al.'s theory and Calderón et al.'s MC simulation in which topological spin defects produce a strong gauge field, proportional to the magnetization and added to the applied physical field. Further an extended scaling hypothesis for the topological spin-defect density explains the behavior of $\rho_{xy}$ along the critical isotherm and beyond. We have argued that the AHE is the first experimental method able to probe the existence and properties of topological spin defects in a 3D Heisenberg-like ferromagnet. The pronounced Berry phase induced by spin texture in a double exchange magnet has opened the opportunity, for the first time, to explore the fundamental role topological defects play in the critical behavior of 3D systems.

**Acknowledgments**

We are grateful to Q. Si, P. Goldbart, Y. Lyanda-Geller, N. Goldenfeld and M.J. Calderón for helpful discussions. HY also thanks T. Banks, T. Namikawa and B. Davidson for their advice about sample growth and micro fabrication. This material is based upon work supported by the U.S. Department of Energy, Division of Materials Sciences under Award No. DEFG02-91ER45439, through the Frederick Seitz Materials Research Laboratory at the University of Illinois at Urbana-Champaign.

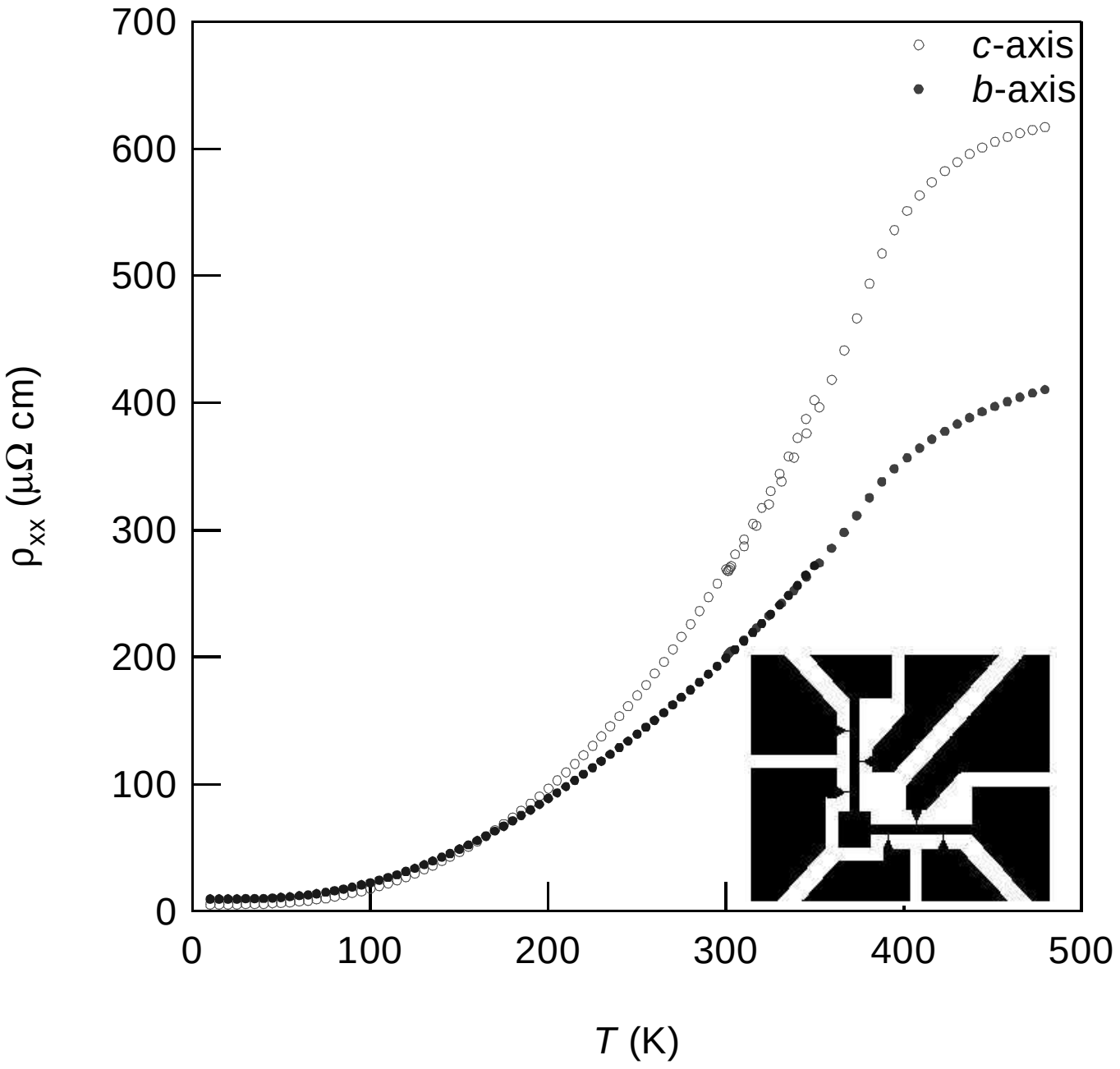

FIG. 1: Resistivity of $CrO_2$ for $b-$ and $c-$ directions as a function of temperature. The $\rho_{300\mathrm{K}}/\rho_{10\mathrm{K}}$ ratio is 20 for $b-$axis and 66 for $c-$axis. The upper inset shows the temperature derivative of $\rho_{xx}$ for both directions. The rounded region is excluded as not in the critical region for fitting. The lower inset is a Hall pattern for this experiment to measure both $b-$ and $c-$ directions at once.

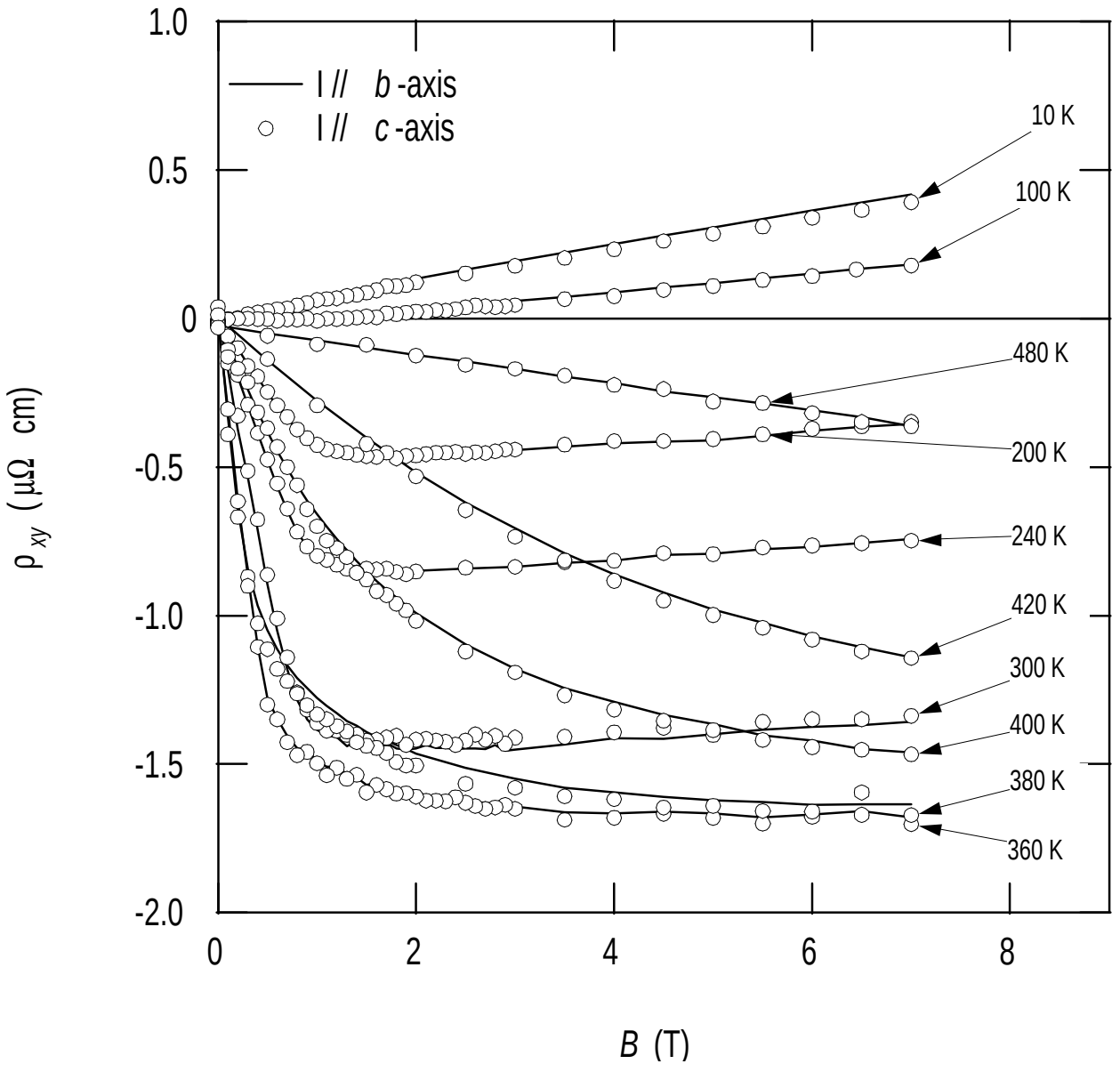

FIG. 2: Field dependence of Hall resistivity $\rho_{xy}(H)$ at various temperatures. At low temperature, the anomalous part cannot be observed but a slight kink is seen at low field, reflecting the complicated multiband structure. Around $T_C$, the anomalous Hall component is dominant and has a sign opposite to the ordinary Hall coefficient. The solid lines and open circles are the Hall resistivity for $b-$ and $c-$ directions, respectively. They are identical within the experimental error.

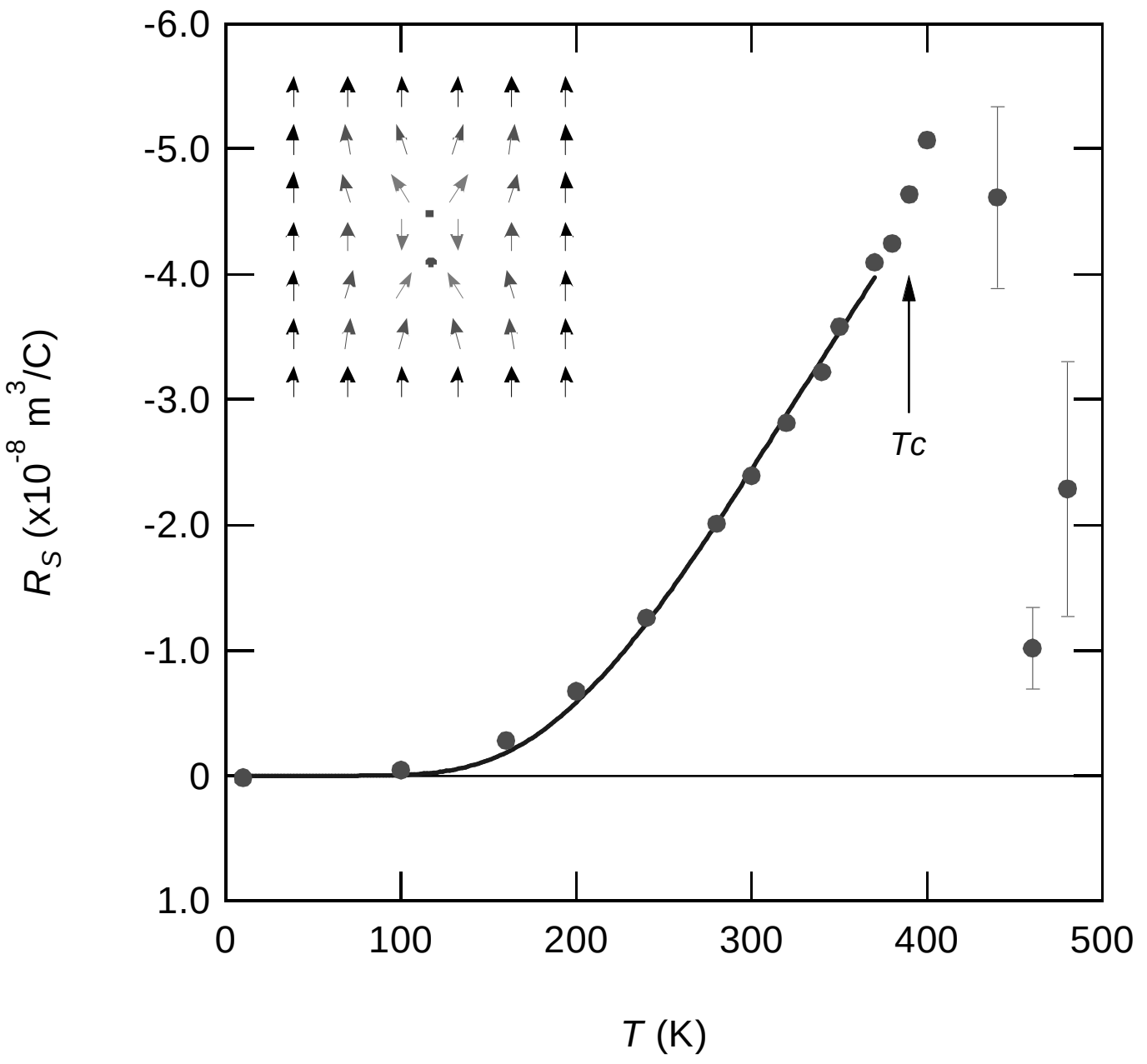

FIG. 3: Anomalous Hall coefficient $R_S$ obtained by fitting of Eq. (1). Because of the difficulty in separating $R_S$ from $R_0$, the data points in the paramagnetic regime are uncertain, the points below 400 K are easily decomposed. The solid line is a fitted result of Eq. (3). $R_S = A\exp(-E_C/k_BT)/T$ is a practical fitting function. The activation energy of the topological defects and a prefactor of the fitting function are obtained as $E_C = 1103 \pm 28$ K and $A = -2.90 \pm 0.25 \times 10^{-4}$ m$^3$K/C, respectively. Inset: Cartoon of a pair of topological spin defects (skyrmion string) with $Q = \pm 1$. The hedgehog and anti-hedgehog are strongly bound, one lattice spacing apart.

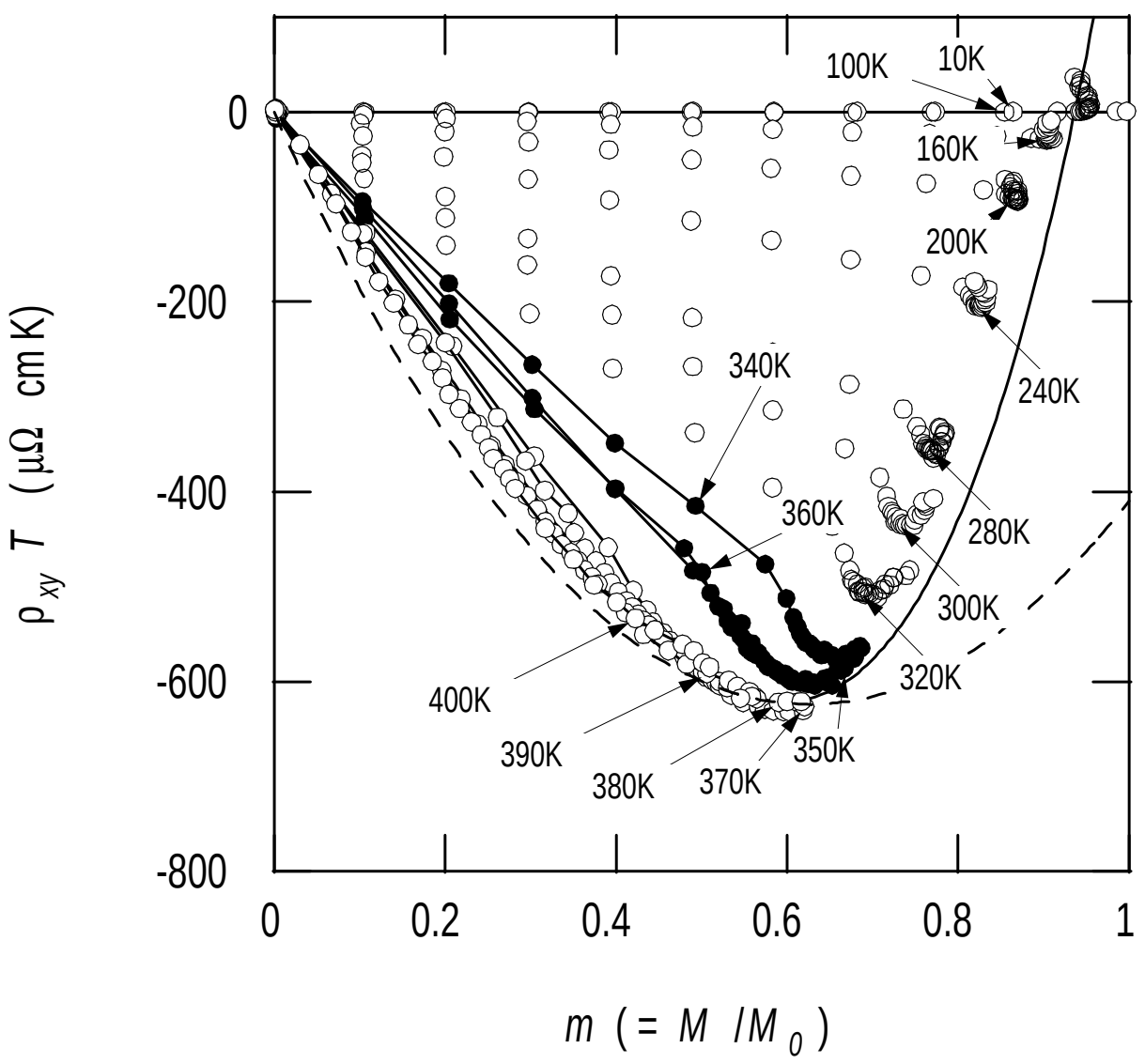

FIG. 4: $\rho_{xy}T$ vs reduced magnetization $m$ at various temperatures. In the critical regime, all points collapse onto a single curve which has an extremum at $m \approx 0.6$. The thick solid line is the best fit for the modified scaling function of the anomalous Hall resistivity, Eq. (6) at the critical isotherm ($x = 0$), and the obtained parameters, $\rho_{xy}^0$ and $D(0)$ are -3.57 $\mu\Omega$ cm, and 1.2, respectively. The dashed curve is the best fit using $\psi = 0.65$ in place of $\alpha$ with $D(0) = 0.8$ and $\rho_{xy}^0 = -5.34$ $\mu\Omega$ cm